\documentclass{article}
\usepackage[T1]{fontenc} 
\usepackage[utf8]{inputenc} 
\usepackage{ismir,amsmath,amssymb,amsfonts,cite,url}
\usepackage{graphicx}
\usepackage{color}
\usepackage{xcolor}
\usepackage{algorithm}
\usepackage{algorithmic}
\newcommand{\CUSTOMCOMM}[1]{\(\triangleright\) #1}
\usepackage{caption}
\usepackage{multirow}
\usepackage{subcaption} 
\usepackage{lineno}
\usepackage{booktabs}
\usepackage[bookmarks=false,hidelinks]{hyperref}
\usepackage{musicography}

\newcommand{\modelone}{$\mathcal{I}$}
\newcommand{\modeltwo}{$\mathcal{F}$}
\newcommand{\algothree}{RefinPaint}

\title{Music Proofreading with RefinPaint: \\ Where and How to Modify Compositions given Context}

\threeauthors
  {Pedro Ramoneda\textsuperscript{\musFlat}\thanks{\hspace{-0.3cm}\musFlat~Work conducted at Sony Computer Science Laboratories, Inc. Tokyo.\vspace{-0.5
  cm}}} {Universtat Pompeu Fabra\\ Barcelona \\ {\tt \small pedro.ramoneda@upf.edu}}
  {Martin Rocamora} {Universtat Pompeu Fabra\\ Barcelona \\ {\tt \small martin.rocamora@upf.edu}}
  {Taketo Akama} {Sony Computer Science Laboratories \\ Tokyo \\ {\tt \small taketo.akama@sony.com}}

\def\authorname{P. Ramoneda, M. Rocamora \& T. Akama}

\begin{document}

\maketitle

\begin{abstract}

Autoregressive generative transformers are key in music generation, producing coherent compositions but facing challenges in human-machine collaboration. 
We propose RefinPaint, an iterative technique that improves the sampling process. It does this by identifying the weaker music elements using a feedback model, which then informs the choices for resampling by an inpainting model.
This dual-focus methodology not only facilitates the machine's ability to improve its automatic inpainting generation through repeated cycles but also offers a valuable tool for humans seeking to refine their compositions with automatic proofreading.  
Experimental results suggest RefinPaint's effectiveness in inpainting and proofreading tasks, demonstrating its value for refining music created by both machines and humans. This approach not only facilitates creativity but also aids amateur composers in improving their work.

\end{abstract}
\section{Introduction}\label{sec:introduction}

Advanced autoregressive models~\cite{vaswani2017attention,radford2019language} have enabled the automatic generation of complex musical performances~\cite{huang2018music,wu2023compose,thickstun2023anticipatory,yu2022museformer,fradet2023byte}. However, while autoregressive models generate music in a strictly forward-moving manner, human composers often follow a more iterative approach, frequently revisiting and refining earlier sections of a piece before proceeding~\cite{collins2011problem,burnard2012musical,Jacob1996Algorithmic}.   
Although there are some iterative methods for music generation~\cite{hadjeres2017deepbach,huang2017counterpoint,chi2020esnet}, there are still areas for improvement in terms of controllability and human-in-the-loop aspects, such as inferring where to modify composition and inpainting capability to enable partial modification.

\begin{figure}[ht!]
    \centering
    \includegraphics[width=0.90\linewidth]{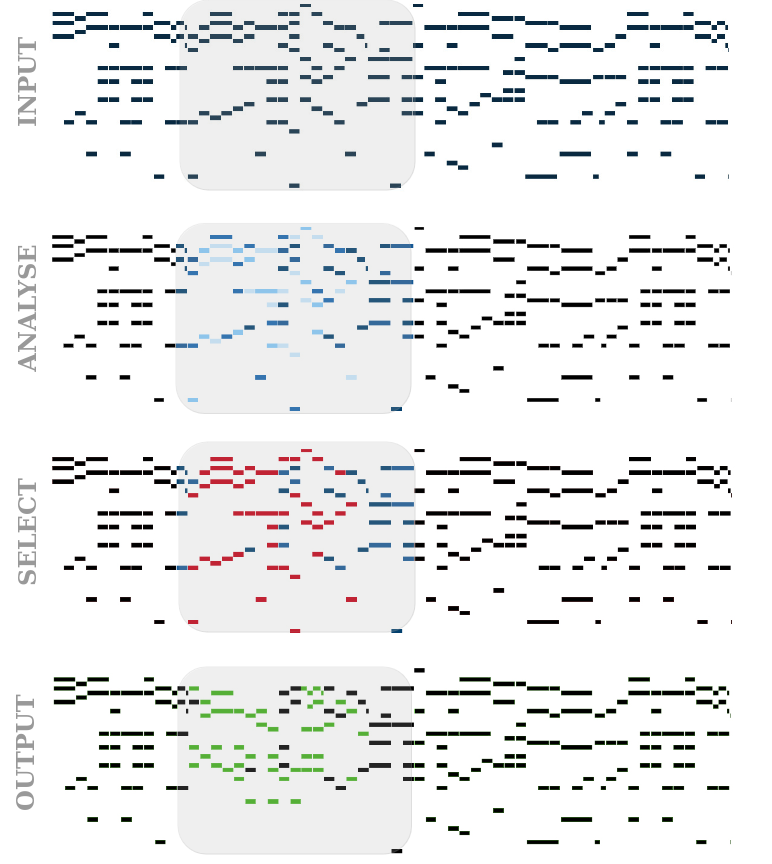}
    \caption{A user selects a MIDI section for enhancement (gray rectangle).  
    Our methodology uses token-level feedback (blue) to highlight critical notes or sequences (red) for regeneration (green). 
    This cycle repeats iteratively.
    }
    \label{fig:teaser}
    \vspace{-0.2cm}
\end{figure}

Iterative refinement proved effective for image generation; in particular, Lezama's Token-Critic~\cite{lezama2022improved} shows how feedback mechanisms can enhance image synthesis. Similarly, such feedback could benefit music composition for iteratively refining generated music.
Within the spectrum of music composition tools, the Piano Inpainting Application (PIA)~\cite{hadjeres2021piano} stands out for its capabilities for automatic music generation that addresses the missing parts of musical performances, a technique referred to as inpainting. We highlight their handling of the musical context both before and after the selected gaps, enabling precise note-level inpainting.
On account of that, inspired by image generation's success with iterative feedback and how PIA handles music context, our research explores applying these concepts to 
enhance controllability, human-in-the-loop functionality, and iterative refinement capability in automatic music generation.

\begin{figure*}[ht!]
\centering
\begin{minipage}{.4\textwidth}
    \centering
    \includegraphics[width=0.6\linewidth]{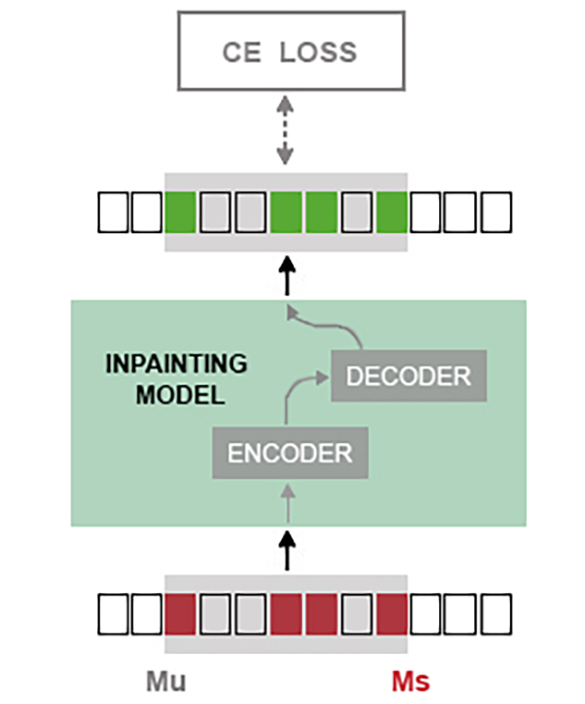}
    \caption{Encoder-decoder architecture for inpainting, given a user-provided mask $M_u$ with a subset mask $M_s$. }
    \label{fig:algo1}
\end{minipage}
\hspace{0.1\textwidth}
\begin{minipage}{.4\textwidth}
    \centering
    \includegraphics[width=0.38\linewidth]{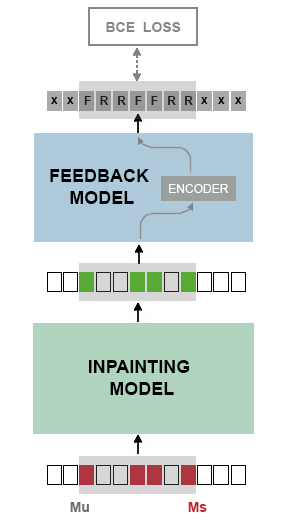}
    \caption{The Feedback algorithm identifies the most realistic tokens by training it to discern between real and synthetic music tokens.}
    \label{fig:algo2}
\end{minipage}
\begin{minipage}{0.90\textwidth}
        \centering
        \includegraphics[width=0.8\linewidth]{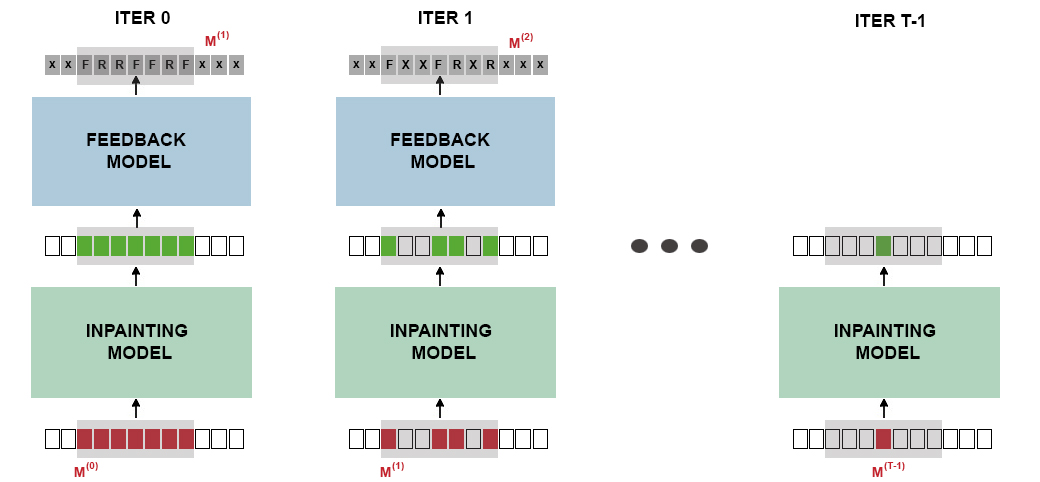}
        \caption{RefinPaint uses inpainting and feedback models to iteratively suggest changes, based on specific note feedback. It reduces the selected tokens in each iteration.}
        \label{fig:algo4}
\end{minipage}
\end{figure*}

In this work, drawing from Token-Critic and PIA, we propose RefinPaint, which aims to boost automatic inpainting and proofreading in music generation. 
Our approach includes an iterative process of identifying areas in a composition needing modification and applying inpainting techniques to these areas. In this context, proofreading refers to automatically identifying and correcting errors or inconsistencies in a music composition. This dual-focus methodology facilitates the machine's ability to improve its automatic inpainting generation through repeated cycles, and offers a valuable tool for humans seeking to refine their compositions with automatic proofreading.

Our RefinPaint method is grounded in an autoregressive inpainting model to generate synthetic music tokens and a feedback model trained to distinguish between original and synthetic tokens. This differentiation is key during the sampling stage when deciding on token retention or revision. RefinPaint takes an iterative approach, integrating feedback into the inpainting model for selectively regenerating parts in each iteration, as Figure~\ref{fig:teaser} shows.
In contrast to Token-Critic, RefinPaint focuses on modifying a specific part of a composition using a contextual model and exposes the intermediate outputs of the autoregressive inpainting model to human inspection in each iteration.

The human-in-the-loop approach we propose allows for selecting the number of tokens to modify
and revise the analysis heatmap at each iteration, as described in the following section. Through experimentation, we confirm RefinPaint's effectiveness in inpainting and proofreading tasks, demonstrating its utility for enhancing music created by both machines and humans. 
Finally, we provide a companion page featuring examples\footnote{At: \url{https://refinpaint.github.io/}} and the code along with the trained models of RefinPaint 
for reproducibility\footnote{At: \url{https://github.com/ta603/RefinPaint}}.

\section{Methodology}

Our proposed methodology employs two models: an inpainting model \modelone{}, and a feedback model  \modeltwo{}, alongside our iterative algorithm \algothree{}. Initially, \modeltwo{} identifies areas within a MIDI file that need improvement based on the specific criteria described in Section~\ref{method:feedback}. It uses a heatmap for detailed MIDI token-level feedback, allowing one to assess the context and relevance of each note in the selected region. 
Then, model \modelone{} can regenerate the selected tokens considering the feedback, as described in Section \ref{method:PIA}. The methodology involves using both models iteratively with \algothree{} and encompasses three main stages: training the inpainting model (Section~\ref{method:algo1}), training the feedback model (Section~\ref{method:algo2}), and finally executing the iterative process for MIDI sequence generation (Section~\ref{method:algo3}).

\subsection{Inpainting model (\modelone{})}
\label{method:PIA}

The inpainting model aims to predict, or fill in, missing parts of a MIDI sequence based on a given mask. We adopt an encoder-decoder architecture for sequence-to-sequence tasks, as shown in Figure~\ref{fig:algo1}, inspired by the PIA study for music generation~\cite{hadjeres2021piano}. This model involves an encoder converting input data into a latent representation and a decoder predicting the final output.

With an anti-causal mask, self-attention within the encoder prevents future data access, while with a causal mask, self-attention within the decoder limits access only to previous data.
With an identity mask, cross-attention enforces positional alignment between the encoder and decoder outputs, which is helpful for aligned sequence tasks.

The attention mechanisms are defined as follows, where \( M_{\text{type}} \) is the mask type (anti-causal, causal, or identity):
\begin{equation}
\text{Attention}(Q, K, V) = \text{softmax}\left(\frac{QK^\top}{\sqrt{d_k}} \odot M_{\text{type}}\right)V.
\end{equation}

This structure enhances the capability of the model to handle bidirectional input-output relationships, essential for inpainting, where future context influences the generation process. Furthermore, we add an extra binary embedding to the encoder input with information about the mask $M_s$--the tokens to regenerate--for the inpainting model.

\subsubsection{Training the Inpainting Model (\modelone{})}
\label{method:algo1}

The training process is outlined in Algorithm~\ref{alg:inpainting}. A batch \(x\) is sampled from the MIDI dataset \(\mathcal{D}\), and a random fragment \(M_u\) is chosen for each sample in \(x\) with a length determined by \(t_1\). It is important to note that $t_1$ refers to the length in terms of the token sequence, rather than the MIDI duration. Consequently, a random mask \(M_s\), with the masking ratio controlled by \(\gamma(t_2)\), is then applied to \(M_u\). The forward pass of the model calculates the loss using the batch \(x\), the mask \(M_s\), and the Cross Entropy (CE) loss function to evaluate the difference between the predicted outputs and the actual labels. The model is subsequently updated via gradient descent. 
The function $\gamma$, a cosine scheduler, dynamically adjusts the masking ratio. It operates on a domain defined by a random variable $t_2$ within the interval $[0, 1]$. Specifically, for any chosen value $t_2$ drawn uniformly from the interval $[0, 1]$, the value undergoes a cosine transformation $\gamma$ to determine the masking ratio, where $\gamma(t_2) = \cos\left(\frac{\pi t_2}{2}\right)$.


\begin{algorithm}
\caption{Training the Inpainting model (\modelone{})}
\label{alg:inpainting}
\begin{algorithmic}[1]
\REQUIRE MIDI dataset $\mathcal{D}$, Inpainting model \modelone{}, ground-truth label $y$
\WHILE{convergence}
    \STATE $x \sim \mathcal{D}$  \hfill\CUSTOMCOMM{Sample batch}
    \STATE $t_1 \sim U(0.1,0.6)$, $t_2 \sim U(0,1)$
    \STATE $M_u \leftarrow \text{Fragment}(t_1)$
    \STATE $M_s \leftarrow \text{RandomMasking}(M_u, \gamma(t_2))$\\
    \STATE $L \leftarrow \text{CE}(\mathcal{I}(x \odot M_s), y \odot M_s)$
    \STATE \text{GradientDescent}(L)
\ENDWHILE
\end{algorithmic}
\end{algorithm}

\vspace{-0.3cm}
\subsection{Feedback model (\modeltwo{})}
\label{method:feedback}

We employ an encoder-only transformer architecture for the feedback phase that classifies music tokens as fake or real. We use this output distribution to select the \(k\) most realistic tokens to retain while the others are regenerated. Unlike the encoder-decoder inpainting model, \modelone{}, this model processes the input through a parallel and bidirectional attention mechanism without employing any attention masks, thus facilitating an unrestricted analysis of the musical context. Additionally, we add an extra binary embedding to the encoder input with information about the mask $M_u$--the selected fragment--for the feedback model.

\subsubsection{Training the Feedback model (\modeltwo{})}
\label{method:algo2}

\begin{algorithm}
\caption{Training the Feedback model (\modeltwo{})}
\label{alg:critic}
\begin{algorithmic}[1]
\REQUIRE MIDI dataset $\mathcal{D}$, Inpainting model \modelone{}, Feedback model \modeltwo{}, Real or fake label creation function $\text{Label}(\cdot)$

\WHILE{convergence}
    \STATE $x \sim \mathcal{D}$ \hfill \CUSTOMCOMM{Sample batch}
    \STATE $t_1 \sim U(0.1,0.6)$, $t_2 \sim U(0,1)$
    \STATE $M_u \leftarrow \text{Fragment}(t_1)$
    \STATE $M_s \leftarrow \text{RandomMasking}(M_u, \gamma (t_2))$
    \STATE $\hat{x} \leftarrow \mathcal{I}(x \odot M_s)$\\
    \STATE $L \leftarrow \text{BCE}(\mathcal{F}(\hat{x}, M_u) \odot M_u, \text{Label}(M_s) \odot M_u)$

    \STATE \text{GradientDescent}(L)
\ENDWHILE
\end{algorithmic}
\end{algorithm}

After training the inpainting model \modelone{}, we train an encoder-only feedback model \modeltwo{}.  This model aims to evaluate the output from \modelone{}, offering feedback on the composition quality of each music fragment denoted by $M_u$.

One ideal way of training \modeltwo{} would involve  a vast dataset of computer- or human-generated music compositions and human experts' revisions for inpainting and proofreading applications. Instead, we propose a more feasible synthetic training strategy, described in Algorithm~\ref{alg:critic}. The inpainting model \modelone{} generates tokens within the selected fragment of a music piece, $M_u$, which we label as `Fake', while we label as `Real' the original unchanged tokens. We utilize these labels to instruct \modeltwo{}, following the process illustrated in Figure~\ref{fig:algo2}.

The training of \modeltwo{} is based on the output of \modelone{}. We begin by sampling a batch $x$ from the dataset $\mathcal{D}$, then apply masking $M_s$ and $M_u$. Model \modelone{} regenerates specific tokens within $x$, yielding a modified output $\hat{x}$. Model \modeltwo{} then assesses each token of $\hat{x}$ against $M_s$, categorizing them as `Real' or `Fake'. The loss $L$ for \modeltwo{} is computed using the Binary Cross Entropy (BCE) loss function, and is minimized through gradient descent. The outcome is a heatmap for $M_u$, which indicates the probability of each token being `Real' or `Fake', determined by the sigmoid activation of the model output.

\vspace{-0.3cm}
\subsection{Generation of MIDI sequences (\algothree{})}
\label{method:algo3}

We capitalize on the strengths of the inpainting and feedback models for the iterative MIDI sequence generation. 
The process shown in Figure~\ref{fig:algo4} begins with a MIDI sequence \(x\) introduced by the user, setting the stage for a loop that spans a predetermined number of iterations \(T\). 

Initially, the user selects the fragment to be modified \(x_m^{(0)}\) and sets the initial selection rate \(k = 0\) for complete inpainting. Alternatively, different values for \(k\) allow the user to control how much of the content to keep in the selected fragment when proofreading.

In the proposed Algorithm~\ref{alg:generation}, at each iteration \(t\), the inpainting model \modelone{} generates a new version of the sequence \(\hat{x}\), based on the current masked input \(x_m^{(t)}\). In the human-in-the-loop scenario, the user can then adjust this generated sequence. The feedback model \modeltwo{} evaluates \(\hat{x}\) and provides a new mask \(M^{(t+1)}\), which the user may also modify. This mask highlights the tokens that are deemed most realistic. The number of selected realistic tokens \(k\) follows a decreasing function \(\gamma\) of the iteration \(t\), which models the increasing confidence in the tokens produced over time. Moreover, we add an extra binary embedding to the encoder input with information about the mask M--the given context--where M changes over iterations.

Refining the music sequence through each iteration aims to achieve a compositional process that closely aligns with that of a human composer so that the user intervention becomes interpretable and natural. It fosters a collaborative environment between the user and the machine and tailors the generation process to the user's specific directives and preferences.


\begin{algorithm}
\caption{Generation Algorithm (\algothree{})}
\label{alg:generation}
\begin{algorithmic}[1]
\REQUIRE Inpainting model \modelone{}, Feedback model \modeltwo{}, masked MIDI $x_m^{(0)}$, No. masked tokens $N$, No. iterations $T$ \\
\FOR{$i = 0$ to $T-1$}
    \STATE \( k = \left\lceil \gamma \left( \frac{i}{T} \right) \cdot N \right\rceil \)
    \STATE $\hat{x} \leftarrow \mathcal{I}(x_m^{(i)})$
    \IF{$i \neq T-1$}
        \STATE $M^{(i+1)} \leftarrow \text{CreateMask}(\mathcal{F}(\hat{x}, M_u))$\\
        \STATE $x_m^{(i+1)} \leftarrow \hat{x} \odot M^{(i+1)}$ with $k$-realistic tokens
    \ENDIF
\ENDFOR
\end{algorithmic}
\end{algorithm}

\vspace{-0.6cm}
\section{Related work}

Automatic music generation has rapidly advanced recently. Significant progress has been made~\cite{wu2023compose,thickstun2023anticipatory,yu2022museformer}, especially in solo piano compositions~\cite{huang2018music,hadjeres2021piano,fradet2023byte}, through the capabilities of autoregressive models in producing coherent musical outputs. However, several challenges remain for creating successful interactions with humans~\cite{hadjeres2021piano,huang2018music,Huang2020,akama2019,akama2020,akama2021,payne2019musenet,hadjeres2017deepbach,hadjeres2020vector,shih2022theme}.

Previous work has explored various approaches to generate music iteratively and allowed for partial modification---often referred to as inpainting---, which enhances controllability. Among them, sequential handling of musical elements has been a common strategy, as in models like DeepBach~\cite{hadjeres2017deepbach} and Coconet~\cite{huang2017counterpoint}. Although these models allow for inpainting and iterative generation, they often rely on random iterations without a mechanism for discriminative feedback to guide improvements. This lack of directed refinement contrasts with the human compositional process, which typically involves iterative improvements based on evaluative feedback. Our proposed approach addresses this limitation by incorporating a feedback model that identifies areas for improvement for both humans and machines to refine the composition.

Although it is not designed as an inpainting model, ES-Net's approach to music generation integrates generative and discriminative capabilities in one model~\cite{chi2020esnet}, with a feature for correcting past errors for iterative refinement. Our model differs significantly: it takes into account the context of the selected fragment, could improve any existing inpainting model, and can handle general MIDI formats.  In \cite{transformer-gan}, the authors propose a GAN model for piano music composition with a discriminator model that discerns real and fake compositions in the training process. However, it does not give feedback on which generated parts are good or bad and does not create compositions iteratively. Yet, the application of discriminative feedback in music generation, particularly in a manner that mimics human iterative refinement, remains largely unexplored.

Finally, inpainting models in music have seen various approaches but remain less studied compared to their counterparts in image generation \cite{elharrouss2020image}. They typically focus on quantized scores, with significant contributions like Gibbs sampling for Bach chorales \cite{hadjeres2017deepbach} and RNN-based melodies inpainting \cite{hadjeres2020anticipation}. 
Studies on transformers for multitrack inpainting have advanced the field, such as MMM~\cite{ens2020mmm}, which utilizes a decoder architecture akin to GPT2~\cite{radford2019language}, and PIA~\cite{hadjeres2021piano}, which uses a specialized transformer design.
We chose PIA over MMM as a ground element in this work, given it is capable of working in the token level or larger contexts and inpainting multiple little fragments at the same time, similar to Token-Critic's generator~\cite{lezama2022improved}.

\vspace{-0.3cm}
\section{Experimental Setup}
\vspace{-0.1cm}

\subsection{Data preparation}
\label{dataset}

Our study utilizes the Lakh MIDI dataset (LMD), an extensive collection of approximately 170k unique multitrack MIDI files, compiled by Colin Raffel for music research~\cite{raffel2016learning}. The dataset offers a wide variety of music, albeit with varying quality due to its internet-sourced nature. Despite this, the volume and diversity of the LMD dataset make it a valuable asset for our proofreading task. We extracted only the piano parts, totaling 120,000 tracks.

We tokenize the piano tracks using REMI (REvamped MIDI-derived events)~\cite{Huang2020}, a music representation method that converts MIDI events into a structured format optimized for Transformer-based models that significantly enhances their ability to comprehend and produce music. REMI categorizes music elements into distinct event types, including timing for rhythm and note events for melody, but we exclude velocity events for simplicity. Specifically, we use a modified version of REMI tailored for handling single-track piano performances, as implemented in~\cite{miditok2021}. The dataset was split into training (hashes 0–d), validation (hash e), and testing (hash f) segments, based on each file's MD5 hash's leading digit, akin to previous methods~\cite{thickstun2023anticipatory,yu2022museformer}

\vspace{-0.2cm}
\subsection{Model development}
\label{sec:modeldev}
\vspace{-0.1cm}

We train the inpainting and feedback models with the AdamW optimizer, using eighty per cent of the dataset for training and the remainder for validation. Each epoch consists of a randomly selected fragment from the training set, 512 tokens in length. We also employ an augmentation procedure that transposes the pitch tokens of a sequence by adding or subtracting up to 6 semitones. For the inpainting model, we apply a cross-entropy loss and use the maximum batch size that our system can handle; a single V100 GPU with 16GB allows for 48 samples. The encoder-decoder inpainting model comprises 12 layers: 4 encoder layers and 8 decoder layers, similar to the original PIA, with 8 heads and an embedding dimension of 512. We employ a cosine scheduler for training, with 16,000 warmup steps, reaching up to a 0.0006 learning rate. The feedback model consists of 6 layers, with an embedding dimension of 512, a dropout rate of 0.1, 8 heads, and the same cosine scheduler. Finally, we acknowledge that optimizing these models was not the main focus of this paper, so there might be better hyperparameter values.

In the particular case of proofreading without human intervention, i.e. for evaluation purposes, the final output is the iteration that maximizes the feedback model probability distribution. Using a sigmoid function, the model determines whether each token in a sequence is fake or real. By averaging the output probabilities, we calculate a global feedback score (GFS) for the sequence's overall realism and select the best regeneration output based on it.

\vspace{-0.3cm}
\section{Inpainting results}

\subsection{Divide and conquer with the inpainting model}

We conducted an experiment to explore how the model's inpainting performance is affected by the percentage of tokens to inpaint in a selected fragment. We hypothesize that the more tokens to inpaint, the harder the problem is, so the model performance is lower. The experiment uses the inpainting model trained as detailed in section~\ref{method:algo1}, and we report its Negative Log-Likelihood (NLL) loss and perplexity of the next predicted token. The evaluation covered the entire test set, with masking ratios ranging from 1 (fully masked) to 0 (no tokens to inpaint) and  a fixed 30\% fragment size rate of the 512 tokens sequence. Results shown in Table~\ref{tab:masking} indicate better performance with reduced masking, confirming our hypothesis. Notably, the average Perplexity value is less than half at 0.05 compared to the 1.0 masking ratio. This finding is crucial for RefinPaint's effectiveness as it reduces the number of tokens to be inpainted in subsequent iterations, considering the iterative process as a top-to-bottom strategy.

\begin{table}[h]
\vspace{-0.2cm}
\centering
\begin{tabular}{ccc}
\toprule
masking ratio & NLL & AVG PPL \\ \midrule 
 0.05 & 0.56 & 0.31\\
 0.10 & 0.58 & 0.33\\
 0.15 & 0.58 & 0.34\\
 0.20 & 0.58 & 0.33\\
 0.40 & 0.64 & 0.41\\
 0.60 & 0.70 & 0.49\\
 0.80 & 0.77 & 0.59\\
 1.00 & 0.86 & 0.73\\
\bottomrule 
\end{tabular}
\caption{Summary of the inpainting experiment with different masking ratios. A masking ratio of 1.0 corresponds to being fully masked, and 0 indicates no masking. The standard deviation is less than 0.01 in all the experiments.}
\label{tab:masking}
\vspace{-0.4cm}
\end{table}

\vspace{-0.35cm}
\subsection{Objective evaluation of proofreading inpainting}
\label{ss:objective}

This section conducts a comparative analysis between the reference inpainting output, as described in~\cite{hadjeres2021piano} (PIA), and our enhanced method. Our method applies the RefinPaint proofreading process to the initial PIA's inpainting output over ten iterations and is referred to as `Ours'. For fragment sizes of 50\%, 30\%, and 10\% of the 512-token test sequences, we computed 1,000 instances each. It is important to note that the PIA method discussed is our reimplementation, since the original code was not available.

Table~\ref{table2} shows the average global feedback score (GFS), computed as explained in Section~\ref{sec:modeldev}, 
and the number of evaluations in which each algorithm outperforms the other (Wins) and in which their scores are the same (Ties).
Table~\ref{table3}, on the other hand, focuses on the comparison between PIA and Ours, employing the NLL loss, a metric of the next token prediction in generated music. This metric, derived from an autoregressive model we trained explicitly from scratch to assess the inpainting results, is a benchmark metric in our evaluation. Similar evaluations have been employed in previous studies in natural language processing~\cite{evaluationloss} and music generation~\cite{roberts2018hierarchical}. Consequently, our study employs a 12-layer Transformer-based autoregressive model with REMI representation. Our goal is to assess the similarity between the distribution of musical elements in inpainted sections and those in the original dataset, including aspects such as rhythms, harmony, or melodies. A lower NLL loss indicates a more accurate prediction of the next token, reflecting a closer approximation to the dataset's inherent musicality. Note we assess this metric over the entire output sequence.

\begin{table}[ht!]
\vspace{-0.1cm}
\centering
\begin{tabular}{cccccccc}
\toprule 
 & \multicolumn{1}{c}{} & \multicolumn{2}{c}{GFS ($\uparrow$)} & \multicolumn{1}{c}{} & \multicolumn{2}{c}{Wins} & Ties \\
\cmidrule{3-4} \cmidrule{6-7}
  & \multicolumn{1}{c}{} & PIA & Ours & \multicolumn{1}{c}{} & PIA & Ours \\
\midrule
50\% & & 0.458 & \textbf{0.696} & & 0 & \textbf{870} & 130 \\
30\% & & 0.515 & \textbf{0.730} & & 0 & \textbf{886} & 114 \\
10\% & & 0.650 & \textbf{0.803} & & 0 & \textbf{891} & 209 \\
\bottomrule 
\end{tabular}
\caption{Comparison of global feedback scores (GFS) between PIA and the proposed RefinPaint methodology, Ours. Higher values indicate better performance.}
\label{table2}

\end{table}

\begin{table}[ht!]
\centering
\vspace{-0.5cm}
\begin{tabular}{cccccccc}
\toprule 
 & \multicolumn{1}{c}{} & \multicolumn{2}{c}{NLL ($\downarrow$)} & \multicolumn{1}{c}{} & \multicolumn{2}{c}{Wins} & Ties \\
\cmidrule{3-4} \cmidrule{6-7}
  & \multicolumn{1}{c}{} & PIA & Ours & \multicolumn{1}{c}{} & PIA & Ours \\
\midrule
50\% & & 2.01 & \textbf{1.97} & & 330 & \textbf{541} & 129 \\
30\% & & 1.68 & \textbf{1.66} & & 347 & \textbf{533} & 120 \\
10\% & & 1.63 & \textbf{1.62} & & 321 & \textbf{457} & 222 \\
\bottomrule 
\end{tabular}
\caption{Comparison of Negative Log Likelihood (NLL) between PIA and the proposed RefinPaint methodology, Ours. Lower values indicate better performance.}
\label{table3}
\vspace{-0.4cm}
\end{table}

Results in Table~\ref{table2} indicate that our model's GFS score is generally better than the baseline, suggesting that the optimization goal of the RefinPaint iterative process is met. The PIA model never wins because this experiment selects the best GFS of all the iterations, as mentioned in Section~\ref{sec:modeldev}. Although dynamic programming or genetic algorithms could enhance the process, this study uses a simpler method, focusing on the iteration with the highest GFS.

In Table~\ref{table3}, RefinPaint consistently achieves a slightly lower average NLL loss than PIA, suggesting that the inpainted content by RefinPaint is more consistent with the original dataset used for training. Furthermore, RefinPaint wins more evaluations than PIA across all the percentages of fragment size evaluated. This further underscores the enhanced performance of RefinPaint in producing sequences more akin to human compositions. However, comparing both tables, we acknowledge that higher GFS does not always imply a better NLL loss, calling for other types of evaluation, as addressed in the next section.

\vspace{-0.2cm}
\subsection{Listening test of proofreading inpainting}
\vspace{-0.2cm}

While computational metrics provide valuable insights into the quality of our inpainted music sections, human perception adds another perspective for evaluating musical quality and appeal. A user-based evaluation was conducted to capture a holistic view of the inpainted outputs' musicality.

For each experiment, which involved 50\%, 30\%, and 10\% fragments of inpainted content, 15 different annotators evaluated both the first iteration of inpainted content (PIA) and the complete iterative process of RefinPaint (Ours) for ten iterations. Participants were exposed to two scenarios, Experiment 1 and Experiment 2: one from the PIA model and one from our RefinPaint model. The order in which these pairs were presented was randomized to avoid any bias. Additionally, we provided the original music fragment without the inpainted content for reference. Participants listened to both the PIA and RefinPaint versions before making their evaluations. They were asked to assess the inpainted content's quality by comparing it to the original fragment, focusing specifically on coherence and creativity. To make their choice, participants were given four options to prevent bias: `Experiment 1,' `Maybe Experiment 1,' `Maybe Experiment 2,' and `Experiment 2'.

\begin{figure}[ht!]
    \centering
    \includegraphics[trim={0.0cm  0.5cm 0cm 0cm},clip, width=0.99\linewidth]{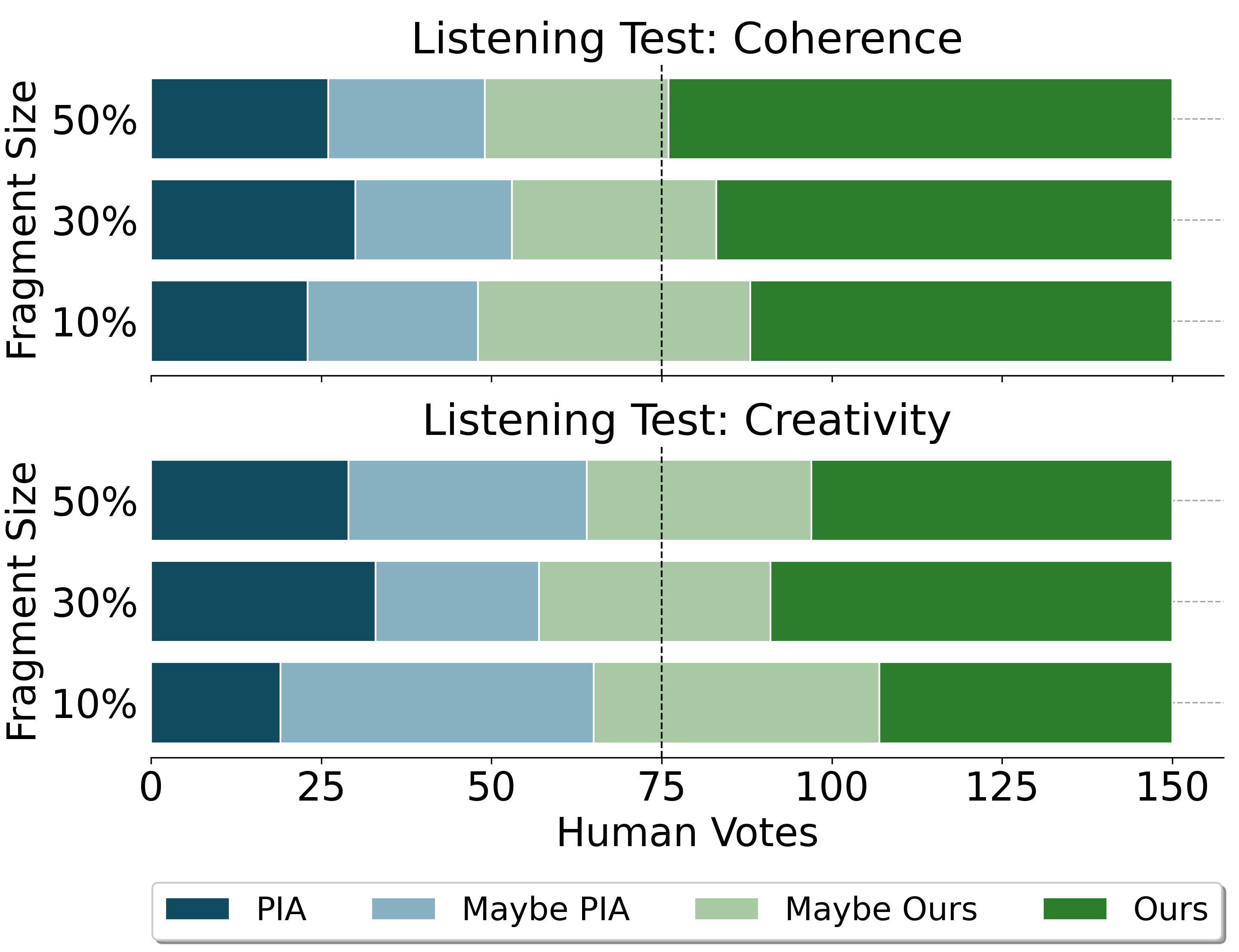}
    \caption{Results of the participants' votes for the listening test comparing PIA and RefinPaint (Ours) along different fragment sizes (50\%, 30\%, and 10\%). } 
    \label{fig:listening}
    \vspace{-0.4cm}
\end{figure}

Figure~\ref{fig:listening} shows the listening test results. Firstly,  PIA got lower preference scores than RefinPaint for the different fragment size conditions. 
In addition, RefinPaint's performance for different fragment sizes shows that the coherence scores increase as the fragment size gets larger, even if the creativity varies. This means that as there is more to inpaint, RefinPaint gets better at being coherent. In contrast, PIA does not show such a strong trend.

The quantitative and qualitative evaluations point towards a clear trend: Refinpaint tends to yield superior inpainting results when proofreading machine inpainted sections compared to the baseline. Our methodology produces music sequences that are more consistent, perceptually closer to the original, and preferred by listeners.

\vspace{-0.1cm}
\section{Case of study on proofreading amateur compositions}
 
We conducted an additional study to explore the proposed system's capabilities for proofreading music compositions by humans. Given the intrinsic difficulties of such a study and due to practical restrictions, we limited our experiment to four amateur composers--two with classical music training and two with modern popular music training. 

Participants used a straightforward proofreading interface that enables bar selection for regeneration, allowing them to choose how much of the content to keep in certain sections of their work, as described in \ref{method:algo3}.  
Additionally, we allowed the users to change the RefinPaint feedback in the selected area and experiment with the tools by conducting as many trials as they wanted.

After testing our inpainting tool on a 30-second music piece, participants responded to questions about their experience. They evaluated whether the tool (i) enhanced their original draft, (ii) sparked new ideas, (iii) could save time over manual proofreading, and (iv) was something they would use in the future. All chose ``yes'' for (i), (iii) and (iv) with three ``yes'' and one ``maybe'' for (ii), suggesting time efficiency as a key advantage and providing an overall positive view of the tool. 

The positive feedback prompted us to showcase the proofread compositions on our companion website. Participants suggested the tool could be particularly effective in overcoming creative blocks, noting that inspiring ideas stemmed from all iterations, not just the last one. Additionally, two participants especially valued the option to alter tokens within the RefinPaint selection.

\vspace{-0.3cm}
\section{Conclusion}
\vspace{-0.2cm}

In conclusion, our novel approach, RefinPaint, significantly enhances music generation by identifying and improving weaker musical elements through iterative feedback. Its effectiveness in both inpainting and proofreading tasks promises a new direction for creative assistance and quality enhancement in compositions by humans and machines alike. Future work could fruitfully extend the research to multitrack compositions and explore control mechanisms for this model, such as conditioning by harmony, rhythm, genre, or other musical factors.

\section{Ethics Statement}

While RefinPaint can represent a significant leap forward in music composition technology, ensuring ethical deployment and use is crucial. We advocate for a future where such technologies support and enrich the creative process, complementing rather than displacing human creativity.
While RefinPaint aims to democratize music creation, making it accessible and achievable for amateurs, there is a risk that professional musicians and composers could feel their roles and contributions are being undermined or replaced by machines. It is essential to strike a balance where this technology serves as a tool for enhancement and learning rather than a substitute for human creativity.
Furthermore, it will be vital to establish guidelines that protect the intellectual property rights of original compositions, whether entirely human-made or AI-assisted.

\bibliography{ISMIRtemplate}

\end{document}